# ALS-U AR RF EQUIPMENT PROTECTION SYSTEM

N. U. Saqib*, C. Toy, Q. Du, K. Bender, S. Basak, S. Murthy, J. H. Lee, D. Nett,
K. Baptiste, Lawrence Berkeley National Laboratory, Berkeley, CA, United States
W. Lewis, Osprey Distributed Control Systems, Ocean City, MD, United States

*Abstract*

This paper presents the design and status of Accumulator Ring (AR) RF Equipment Protection System (EPS) of Advanced Light Source Upgrade project at LBNL. The key components of AR RF EPS include a Master Interlock PLC subsystem handling supervisory control and slow interlocks in ms scale, an FPGA-based LLRF Controller managing fast interlocks in µs scale, a 60 kW high-power amplifier with standalone PLC-based slow (ms scale) and FPGA-based fast (µs scale) protection systems, and an RF Drive Control Chassis acting as primary RF mitigation device. The design of AR RF EPS is presented along with internal RF and external AR subsystems interfaces.

## INTRODUCTION

The Advanced Light Source (ALS) at Lawrence Berkeley National Laboratory, a U.S. Department of Energy synchrotron light source user facility operational since 1993, features a 196.8 m Storage Ring (SR) supporting multi-bunch operations with a 500 mA electron beam, delivering synchrotron X-rays to 40 beamlines and end-stations. An upgrade, Advanced Light Source Upgrade (ALS-U), is underway to increase soft X-ray brightness and flux at 1 keV by two orders of magnitude, reaching the diffraction limit. Furthermore, ALS-U will offer infrared and hard X-ray capabilities comparable to the existing ALS. To accomplish this, the existing triple-bend-achromat 1.9 GeV SR will be replaced with a new nine-bend-achromat 2.0 GeV SR, and a new full-energy 2.0 GeV Accumulator Ring (AR) will be installed alongside the SR in the same tunnel (Fig. 1). The AR is designed for full-energy swap-out injection and bunch train recovery, and adopts a triple-bend-achromat lattice similar to the current ALS SR. Table 1 outlines the RF specifications of the AR lattice. The AR RF system features two 500 MHz normal-conducting cavities. To provide RF power to each cavity, a chain of RF components is designed comprising of an LLRF Controller, 60 kW solid-state High Power Amplifier (HPA), high-power RF circulator, high-power RF switch, and a rigid RF coaxial transmission line.

For protection of RF components of the two cavities, an Equipment Protection System (EPS) is designed to turn off the high-power RF in the AR upon detecting anomalies. The block diagram of AR RF EPS is shown in Fig. 2. The EPS monitors and interfaces with subsystems of the AR RF, along with the interfaces to external AR subsystems such as Machine Protection System (MPS), Personnel Protection System (PPS) and Vacuum system. The AR RF EPS interlocks the RF drive signal from LLRF controller to the HPA in case a fault occurs. The AR RF EPS consists of a Master Interlock PLC subsystem, two LLRF Controllers, two HPAs and two RF Drive Control Chassis. The primary method to cut off the RF path is to open pin diode having response time of 800 ns situated inside RF Drive Control Chassis, while the secondary approach relies on the exchange of permit/inhibit signals among Master Interlock PLC, LLRF Controllers and HPAs as shown in Fig. 2. Table 2 lists the response time requirements for AR RF shutoff.

---
* nusaqib@lbl.gov

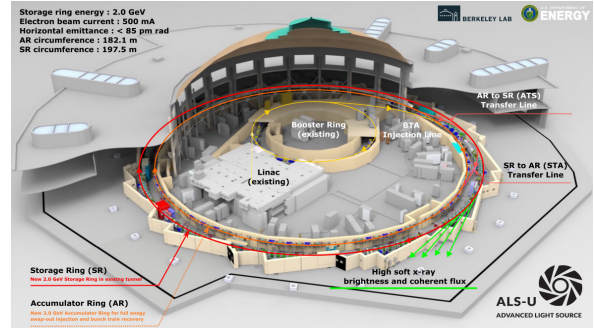

Figure 1: ALS-U.

Table 1: AR RF Specifications

| Specification | Value |
|---|---|
| Cavity RF Frequency | 500.394 MHz |
| Number of Cavities | 2 |
| R/Q (each) | 3.5 |
| Cavity voltage | 500 kV |
| $\beta$ | 1.18 |
| Energy loss per turn | 270 keV |
| BM Beam Power | 13.3 kW |
| Parasitic Beam Power | 0.2 kW |
| Total Beam Power | 13.5 kW |
| Cavity Power (no beam) | 36.0 kW |
| Cavity Power (beam) | 42.7 kW |
| High Power Amplifier | 60 kW |

The next section describes the AR RF EPS components and interfaces.

## DESIGN

The Master Interlock PLC subsystem serves as brain for the AR RF EPS system, providing an array of both interlock and non-interlock functionalities. This includes slow interlock mechanisms, supervisory control and monitoring, coordination of AR RF startup/shutdown, RF cavity tuning, and control of auxilliary devices. The Master Interlock PLC subsystem consists of two PLC chassis: the Master Interlock

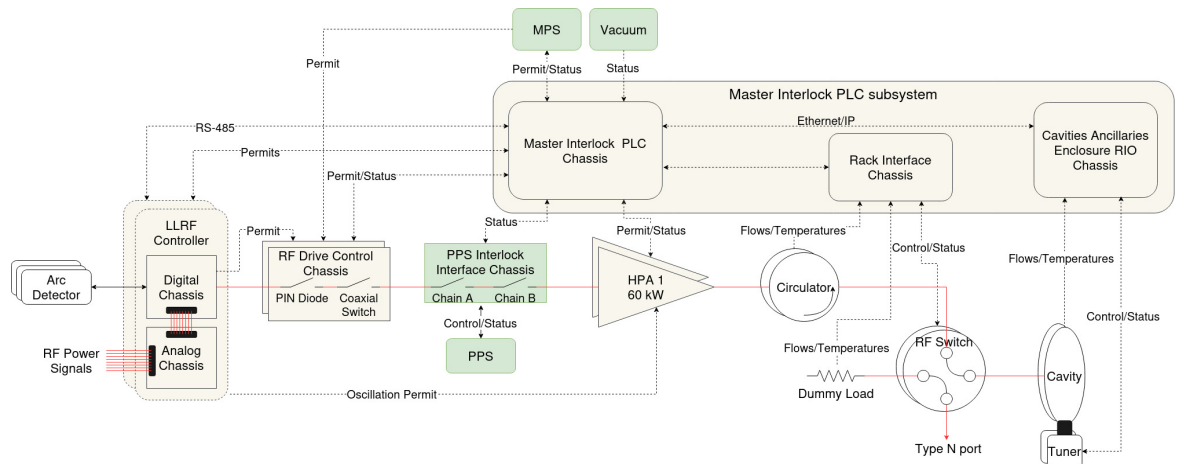

Figure 2: AR RF EPS block diagram.

Table 2: AR RF EPS Response Time Requirements

| Specification | Threshold |
| --- | --- |
| LLRF interlock response | 6 µs |
| Master Interlock PLC interlock response | 10 ms |
| HPA fast interlock response | 10 µs |
| HPA slow interlock response | 800 ms |
| MPS interlock response | 20 ms |

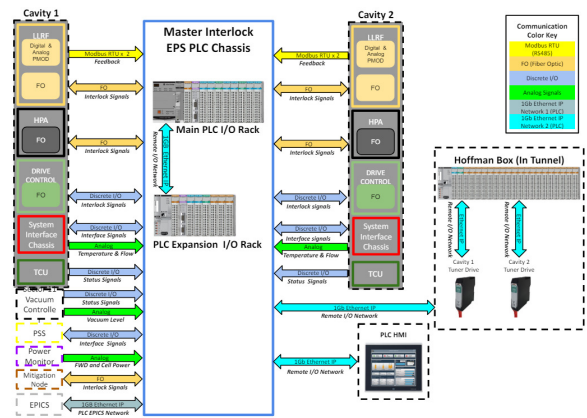

Figure 3: Master interlock PLC subsystem architecture.

PLC Chassis located inside the AR RF rack and the Cavities Ancillaries Enclosure (CAE) remote I/O chassis located inside the AR tunnel. The CAE remote I/O chassis interfaces the two cavities' temperature sensors, flow sensors and stepper motor drive controllers for cavity tuners. The Master Interlock PLC Chassis interfaces with various RF subsystems including LLRF Controllers, HPAs, RF distribution components and external AR subsystems including MPS, PPS and Vacuum. Allen Bradley CompactLogix 5000 series CPU and I/O modules are selected for PLC hardware, and a PanelView 5510 HMI is chosen for local control of the Master Interlock PLC subsystem. Remote control is facilitated through EPICS (Experimental Physics and Industrial Control System) and Phoebus CS-Studio. The Master Interlock PLC hardware architecture is shown in Fig. 3.

The Master Interlock PLC interfaces with external AR subsystems such as MPS, PPS and vacuum. The AR MPS collects permit/inhibit signals from AR subsystems including all sectors Vacuum, Magnets, Power Supplies, Beam Diagnostics and Injection/Extraction and sends redundant permit/inhibit signals to the Master Interlock PLC and RF Drive Control Chassis to turn off the RF in case a subsystem fault occurs that requires global RF mitigation action. The PPS interface is accomplished via PPS Interlock Interface Chassis, which consists of two PPS controlled coaxial switches in series for each AR RF cavity. The chassis provides RF mode selection via safety-system-compatible key-switch system, allowing operators to select one of four operating modes: Operational, RF Test, RF Test with Access and RF Test to Dummy Load. The Master Interlock PLC also interfaces with vacuum gauge controllers of each cavity to ensure proper vacuum environment and interlocks RF drive signal in case of a vacuum breakdown.

The two AR RF solid-state HPA [1] provide stable 60 kW RF power to each cavity at 500.394 MHz, generating the cavity voltage essential for the AR electron beam. The required RF power for each AR cavity is set at 48 kW and to address potential failures and ensure operation under demanding conditions, a 12 kW power margin is maintained, resulting in a maximum HPA output power of 60 kW at the required 500.394 MHz frequency. The HPA features standalone protection system consisting of Allen Bradley CompactLogix 5000 series PLC CPU, a main FPGA, and 55 sub-FPGAs within its modules and subassemblies. It utilizes two categories of interlock response speeds: fast (< 10 µs) for FPGA-based interlocks and slow (< 800 ms) for PLC-based interlocks. The HPA protection system provides comprehensive monitoring and control of all the signals and parameters. Each HPA can be operated locally via dedicated Panel View 5510 HMI or remotely through EPICS and Phoebus CS-Studio. The HPA sends fault status and receives permit signal to/from Master Interlock PLC.

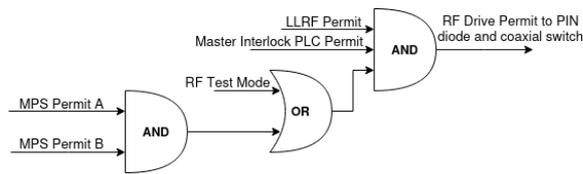

Figure 4: RF drive control chassis hardware logic.

The AR LLRF subsystem [2] comprises of two LLRF Controllers, each responsible for providing RF drive signal to the HPA. The EPS functionality of LLRF controller includes fast RF power and arc detection interlocks. Each LLRF Controller consists of two main components: an Analog Chassis and a Digital Chassis. The arc detectors for circulator, circulator load and a future spare are interfaced with digital chassis. The digital chassis sends a permit signal to the RF Drive Control Chassis and the Master Interlock PLC in the event of arcing or in case RF powers surpass predefined thresholds, and receivers a permit signal from Master Interlock PLC. An Oscillation Permit signal is sent to the HPA to interlock RF drive signal when power discrepancies occur between measured cavity power and the drive signal. Digital and Analog chassis are also interfaced with Master Interlock PLC via RS-485 interface to exchange status and configuration parameters.

RF drive control chassis is the primary mitigation device designed to interrupt the RF drive signal path from LLRF to HPA. The mitigation mechanism involves a PIN diode and an RF coaxial switch, positioned in series within each RF drive control chassis. The switching times of selected RF PIN diode and coaxial switch are 800 ns and 20 ms respectively. The chassis receives permit signals from both the AR RF EPS components including LLRF Controller and Master Interlock PLC, as well as external subsystems such as MPS. The decision to open/close the PIN diode and coaxial switch is governed by the hardware logic illustrated in Fig. 4. The hardware logic allows bypassing the MPS permit signals during AR RF test modes, facilitating commissioning and troubleshooting. The test modes are selected from keyswitches located inside PPS Interlock Interface Chassis.

## DEVELOPMENT STATUS

The AR RF EPS design has reached maturity and fabrication of key components is nearing completion including Master Interlock PLC Chassis, RF Drive Control Chassis, LLRF Controllers, and CAE remote I/O chassis. Software development for the Master Interlock PLC subsystem including PLC code, HMI screens, EPICS IOC, and Phoebus OPIs is in final stages and being tested via simulation. LLRF Controllers firmware, software and EPICS IOC are also being tested extensively. Following the fabrication process, each chassis will undergo bench testing before being installed into one of the three AR RF racks on ALS Booster Ring roof. The two HPAs were manufactured at R&K Japan and delivered at LBNL in December 2023. After having AR RF EPS components installed in place, integrated testing and commisionning is expected to follow in mid-2025. Fig. 5 and Fig. 6 illustrate the recent lab setup and master interlock HMI screen respectively.

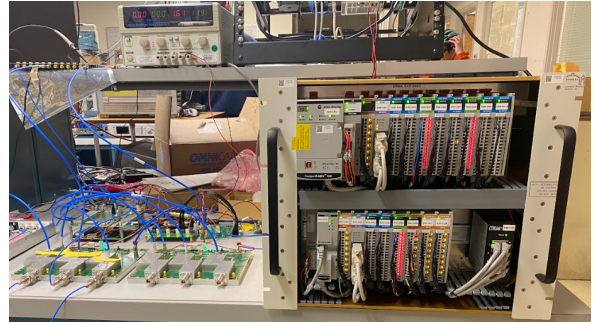

Figure 5: Lab setup - LLRF controller and master interlock PLC chassis.

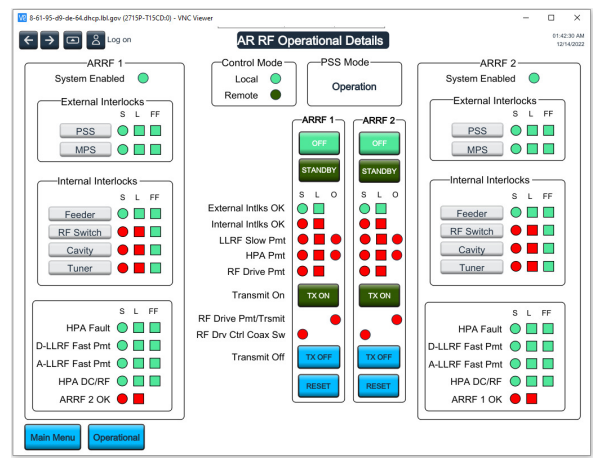

Figure 6: Master interlock HMI - interlocks summary.

## ACKNOWLEDGEMENT

This work is supported by the Office of Science, Office of Basic Energy Sciences, of the U.S. Department of Energy under Contract No. DE-AC02-05CH11231.